\documentclass[prd,preprint]{revtex4-1} 

\usepackage{amsmath}
\usepackage{amssymb}
\usepackage{amsfonts}
\usepackage{graphicx}

\usepackage{verbatim}

\begin{document}

\title{Existence of Minkowski space}

\author{Serge A. Wagner}
\email{s\_wagner@mail.ru}
\noaffiliation

\date{\today}

\begin{abstract}
Minkowski space serves as a framework for the theoretical constructions that deal with manifestations of relativistic effects in physical phenomena. But neither Minkowski himself nor the subsequent developers of the relativity theory have provided a reasonable rationale for this mathematical construct. In physics, such a rationale should show lower-level statements that determine where the proposed mathematical  structure is applicable and yield formal premises for proving its existence.

The above failure has apparently been due to the features of the adopted formalism based on the unjustifiably exclusive use of coordinates in the theoretical analysis of physical phenomena, which ignores the necessity of having physical grounds for mathematical concepts. In particular, the use of a coordinate transformation between two inertial reference frames makes the consideration so cumbersome that it appears useless for solving the fundamental problems of physical theory, including the question of whether Minkowski space exists.

In contrast, a straightforward calculation proves that the transformation of the time and the position vector of a physical event between two physical spaces establishes an equivalence relation between pairs made of these variables. This means the existence of Minkowski space and shows that the premises for its proof are the same as for the coordinate-free derivation of basic effects of the special relativity theory: the use of Einsteinian time variable and motions of particles able to interact with each other and electromagnetic field over a short spatial range only.

The high degeneracy of free motions of point particles, together with the intricacy of the above mentioned calculation, suggests that a further generalization of Minkowski space is beyond belief, so that the modification or even the abandonment of the concept of spacetime seems quite natural.
\end{abstract}


\maketitle

\section{Introduction}\label{introduction}
Physics and relativity textbooks\cite{textbooks} (in agreement with the mathematics monographs\cite{Naber2012,Schroeter2017}) present Minkowski space as a four-dimensional vector space where a system of four coordinates \(t,x,y,z\) is supposed to represent an inertial reference frame with its clock readings \(t\) and spatial Cartesian coordinates \(x,y,z\) so that the quadratic form
\begin{equation}\label{interval squared}
 c^2t^2-x^2-y^2-z^2
\end{equation}
is invariant with respect to a changeover from one coordinate system to another. Here and throughout the article \(c\) is the speed of light.
\subsection{Inception of formalism for special relativity theory}\label{Poincare}
The story began with the report\cite{Poincare1906}, where H.~Poincar\'e had presented the change of time variable and Cartesian spatial coordinates that does not alter the appearance of Maxwell's equations. He had given it the name ''Lorentz transformation'' since in the physical interpretation of this change of the independent variables as a motion of a coordinate system H.~Poincar\'e followed the preceding work\cite{Lorentz1904} of H.~A.~Lorentz, who attempted to explain why the Earth's motion is not detectable with an aid of optical experiments conducted on the Earth's surface.
Only the part of Maxwell's equations that does not involve electric charges appears sufficient in Ref.~\onlinecite{Poincare1906} for inferring Lorentz transformation while in order to obtain the associated transformation of an electric charge density H. Poincar\'e had exploited an implicit assumption that the total charge of a moving charged body (referred to as 'the electron') is independent of the state of the body's motion.

It should be noted that the term ``transformation" in Ref.~\onlinecite{Poincare1906} bears no relation to any specific occurrence among physical bodies but means that, in general, a change of independent variables along with an induced (i.e. appropriately corresponding) change of unknown functions in a set of partial differential equations is expected to modify the appearance of the equations. Thus, one can say that Ref.~\onlinecite{Poincare1906} presents the change of variables so that the transformed equations appear to be the same combination of the transformed electromagnetic field quantities as the source equations with the source electromagnetic field quantities are.
\subsection{Formulation of special relativity theory}\label{Einstein}
In contrast to the analysis of H.~Poincar\'e, A.~Einstein, in his initial defining article\cite{Einstein1905} on the relativity theory, has addressed physical events, such as an interception of small (parts of) physical bodies by light rays/fronts, from the outset. He distinguished a ``stationary system" as a coordinate system where Newtonian equations hold and the measurements are based on the use of a measuring rod and Euclidean geometry.

It should be pointed out that A.~Einstein has described a coordinate system as relations between rigid bodies\cite[S.~892]{Einstein1905} and associated the ``stationary system" with a ``stationary" space\cite[S.~897]{Einstein1905}, so his ``stationary system" is a physical object which this article further refers to as ``an inertial reference frame" to distinguish it from the mathematical concept ``coordinate system"\cite{reference frame}.
(For the refinement of the concept ``space" see Section \ref{physical spaces} below.)

With referring to uniform motion in the ``stationary" space A.~Einstein has also identified a ``moving system"  but then, in accordance with his formulation of the relativity principle\cite{frames' equivalence}, considered it to be on a par with a ``stationary system" so that relations between the two inertial reference frames have turned out to be mutual\cite[S.~903]{Einstein1905}.

In addition, A.~Einstein has extended the concept of the time variable with an aid of the propagation of light (see Section \ref{Einsteinian time} for details.) As a result, he has obtained the formulas identical to those of Lorentz transformation\cite{Lorentz transformation}, which, however, connects the variables that relate to a given event in two inertial reference frames. So, in theoretical physics, one uses the word ``transformation" for the change of the description of a frame-independent object/concept owing to a changeover from one inertial reference frame to another.

In the introductory part of his article\cite[S.~891-892]{Einstein1905} A.~Einstein has extended the principle of the relativity to electrodynamics and, in \S 3 of his article, applied it to Maxwell's equations for the electromagnetic field free of electrical charges (i.e., similarly to Ref.~\onlinecite{Poincare1906}, required that they retain their form when one inertial reference frame is replaced by another.) This has allowed him to obtain the relation between the sets of the electromagnetic field quantities in two inertial reference frames. Then he has assumed it valid in the presence of electric charges and (in contrast to Ref.~\onlinecite{Poincare1906}) has thereby arrived at the transformation law of an electric charge density and the conclusion that the total charge of a charged body is a frame-independent quantity.\cite{quantum}
\subsection{Minkowski's conjecture}\label{Minkowski}
Considering the covariance of Maxwell's equations with respect to Lorentz transformation, in his lecture\cite{Minkowski1908} H.~Minkowski has proposed the technique of those combinations of (as well as relationships between) mechanical and electromagnetic quantities which make it easily perceivable that they preserve their appearance when subjected to Lorentz transformation along with the transformations induced by Lorentz transformation. Apart from these covariant combinations, lately introduced into theoretical physics by Part II of  Ref.~\onlinecite{Einstein&Grosmann1913} as 4-tensors, he has also found some 4-scalars (i.e. the invariants), including the kinematic quantity \eqref{interval squared} and two electromagnetic quantities\cite[S.~68]{Minkowski1908}.

For no reason other than the formal similarity between Lorentz transformation and the transformation equations relating two sets of coordinates of a given spatial point in two Cartesian coordinate systems that differ in the directions of their coordinate axes\cite{Einstein's acknowledge}, H.~Minkowski has called 4-tuple \(x,y,z,t\) a space-time point\cite{worldpoint}. He undeniably implied that, similarly to the relation between different 3-tuples \(x,y,z\) and one spatial point they can represent, there must be an entity that corresponds to a collection of 4-tuples \(x,y,z,t\) connected by Lorentz transformations.

An analogy, however, is not proof and may mislead those who rely on its consequences without regard for its premises. In fact,  even in H.~Minkowski's time some mathematicians did try to construct Euclidean geometry on the basis of motions of rigid forms\cite{Euclid axiomatics}, but obviously H.~Minkowski was hardly interested in their work and was not going to make sure that there was no problem in modifying it for a 4D space equipped with the ``metric" \eqref{interval squared}.

In \S 6 of his lecture, for events at two space-time points in a given coordinate system H.~Minkowski has found the Lorentz transformation to the system where these events occur simultaneous. The reader of that paragraph may think that H.~Minkowski has identified his space-time points with physical events\cite[S.~69]{Minkowski1908} or, in an attempt at following Ref.~\onlinecite{Einstein1905}, has at least mapped events into a set of space-time points, though carelessly mapping a set of one origin/structure into that of another does not necessarily result in a one-to-one correspondence.

Actually, in his approach to space and time concepts H.~Minkowski was only able to refer to human experience\cite[S.~69]{Minkowski1909} while A.~Einstein could write about a \textit{position} of an elementary physical event because the latter, such as detecting a particle or emitting/absorbing light by an atom, could include \textit{interaction} with a small part of that association of solids which represents a body of an inertial reference frame.
\subsection{Einsteinian time variable}\label{Einsteinian time}
The regularities of classical mechanics provide their own possibilities for formulating a definition of the time variable. In particular, as soon as one identifies all physically independent quantities of non-relativistic macroscopic mechanics, usually referred to as mechanical state variables, one sees that the sequence of changes in these quantities is described by an autonomous system of first-order differential equations, derivable from Newton's laws. Then Newtonian time variable arises as the independent variable of this system of equations. Such a definition suggests that, in agreement with human everyday activities, a time variable is not a physical quantity but an auxiliary quantity that facilitates theoretical analysis and other human activities in physics. This definition can be extended to encompass other slow phenomena of macroscopic physics, including electric and magnetic ones, as soon as they appear in the above form.

The generalization of the above definition so as to include relativistic mechanics is not possible, since the laws of relativistic mechanics, such as those describing the interaction of relativistically moving charged particles, have not been known yet. In Ref.~\onlinecite{Einstein1905} A.~Einstein has addressed the lower-level, generally pre-numerical, description/definition of \textit{time moments} as a division of all events into the groups of events observed at the same time moments.

In order to partition\cite[p.~18]{Gallian2010} a set indirectly, mathematics suggests exploiting an equivalence relation\cite[p.~16]{Gallian2010} between each two elements of the set. In Ref.~\onlinecite{Einstein1905} A.~Einstein has called the required relation between two events a synchronization and given a formulation for its properties equivalent to its symmetry and transitivity. In other words, A.~Einstein has essentially demanded that the synchronization of two events be an equivalence relation between them. He has proposed the physical realization for the synchronization between the readings of two clocks in different places with a round-trip of a light pulse between these clocks and adjusting their readings so as to maintain a certain relationship between the times of emitting, reflecting and absorbing the light pulse. In addition, he has given a formula for the speed of light through the size of the round-trip of the light pulse and the times of its emission and absorption and has stated that the speed of light is a universal constant.

The synchronization in the above-described lower-level definition is its only feature specific for the defined quantity. Hereafter in this article, the quantity defined with aid of Einstein's synchronization is referred to as Einsteinian time variable.

The degree to which the theory of relativity is consistent with the already proven consistent description of physical phenomena determines both its area of applicability as a whole and the areas of parameters where its individual concepts are applicable. In particular, it may turn out that Einsteinian time variable is possible only along the trajectory of a point particle. In order to associate Einsteinian time variable with the entire space of possible positions of all particles in a given inertial reference frame, one should at least check transitivity of the synchronization for three space points that do not necessarily belong to any trajectory of any point particle. This article accepts this association with the space (or, equivalently, with all inertial reference frames resting there) as a premise to show that even such a strong assumption brings little to generalize the idea of spacetime after the existence of Minkowski space has been established.

Later, to present a formal derivation of the Lorentz transformation without addressing the definition of time directly, A.~Einstein defined the time of a reference frame as an aggregate of readings of all clocks resting there, with clocks synchronized so that the speed of light appears a universal constant.\cite{Einstein1907b} The renewed definition tacitly implies that reducing the distance between clocks down to their merger affect neither clocks themselves nor their synchronization. Macroscopic clocks with such properties can easily be imagined, but the atoms chosen as sources of standard frequency cannot be such ``clocks", because their interaction must influence them obviously. The less restrictive previous presentation of the relativity theory also helps little, since the relativistic interaction of charged particles has not yet been described.

In Ref.~\onlinecite{Einstein1907b} A.~Einstein has also identified principle of relativity and ``principle of the constancy of the speed of light" as two premises of the above simplified version of his theory. Subsequent authors referred to them as "postulates" and even considered the second ``postulate" worth for experimental tests, ignoring that actually it is a part of the definition of time, made consistent with the principle of relativity. (The fact that the speed of light does not depend on its source relates to the nature of light rather than to Einstein's theory). In Ref.~\onlinecite{Alvager1964} reporting testing the second ``postulate", the characteristic scale of the measured interval has appeared the same as that of Newtonian time so the result has hardly related to special relativity.
\subsection{Post-Einsteinian attitude to Minkowski space}
Although A.~Einstein has acknowledged the ideas of H.~Minkowski as early as in Ref.~\onlinecite{Einstein1907b}, the other physicists showed more caution. They took the ideas of H.~Minkowski as ``a remarkable and instructive graphic representation of the Lorentz transformation"\cite[p.~129]{Silberstein1914}, ``a four-dimensional method of expressing the results of the Einstein theory of relativity"\cite{Tolman1917} and viewed ``absolute world" in Ref.~\onlinecite{Minkowski1909} as a room for mathematical objects invariant with respect to Lorentz transformations (along with transformations induced by Lorentz ones) where there is ``a natural generalization of the ordinary vector and tensor calculus for a four-dimensional manifold."\cite[p.~22]{Pauli1958} In contrast, the authors with no background in usual, non-relativistic, physics accepted H.~Minkowski's ``world" outright and much more enthusiastically.\cite{mathematicians}

The post-Einsteinian generation of physicists got more focused on applications than attentive to foundations. In addition, the authors of textbooks on relativity as well as the lecturers of appropriate theoretical physics courses found it easier to deduce the theory of the relativity from the idea of spacetime. In line with this education practice, scientific authors got their theoretical constructions developed in Minkowski space at the outset when expected them to apply to the manifestation of relativistic effects.

The reasonable estimation of the validity of a theoretical consideration need addressing its premises, based either on well-tested lower-level theoretical relations or directly on experimental data. However, since the proof of existence of Minkowski space has not yet been published up to the present time, the most important part of premises for each of the many theoretical constructions remains unrecognized.

The next section draws the reader's attention to the coordinate-free formalism in the special relativity theory and explains why it is necessary for the required proof. The section \ref{transformation} contains the transformation of the time and the position vector of a point physical event between two spaces. Section \ref{proof} exploits this transformation to make the formal proof of the existence of Minkowski space. Its significance for physical theory is discussed in Section~\ref{discussion}.

\section{Coordinate-free formalism in relativity theory}
\subsection{Physical spaces in mechanics}\label{physical spaces}
It is relations between parts of a solid that underlie the group of motions of rigid bodies. This group includes spatial translations \(\hat T\) and rotations \(\hat R\). The additive representation of a spatial translation is usually referred to as a spatial vector. One can use rotations to introduce an angle between two vectors etc.

When the orthonormal vectors \(\mathbf{e}_x, \mathbf{e}_y, \mathbf{e}_z\) represent the translations along three mutually perpendicular directions, the decomposition
\begin{equation}\label{position vector decomposition}
\Delta\mathbf{r}=\Delta x\,\mathbf{e}_x+\Delta y\,\mathbf{e}_y+\Delta z\,\mathbf{e}_z
\end{equation}
of a displacement (the change of a position vector \(\mathbf{r}\)) is just what defines Cartesian coordinates of a small body that can manifest itself by interaction with a solid. If one invokes some other properties of physical bodies and their motion one finds that the positions of the small body obey Euclidean geometry so that the set of small parts of a conjectural boundless solid can represent a Euclidean space.

The above described physical realization of Euclidean geometry evidently breaks down at sufficiently small scales, where the atomic/molecular structure of any solid is important. In order to extend the validity of a Euclidean space to smaller scales one has no choice but to address Newtonian mechanics of stable charged particles. Since the limits of applicability of Euclidean geometry are not among the topics of this article, the reader may simply accept the assumption that the set of motions of interacting charged particles is rich enough to ensure the existence of angles and other geometric features, including a position vector \(\mathbf{r}\).

Only a couple of additional comments are required here.

To build Euclidean geometry with motions of charged particles one has no way but to exploit lower-level \textit{relations} between their trajectories, such as an interception in a collision of two relatively fast particles, or \textit{internal properties} of a specific trajectory, such as those of a closed orbit of each of two relatively slow oppositely charged particles. Generally, it can result in the constructions of a moving Euclidean space. Then the indispensable step should be such change of \(\mathbf{r}\) that stops the center of mass of the particles.

Thus, a set of non-relativistic communicating observers is always capable to label the positions of particles by position vectors \(\mathbf{r}\) defined in its stationary Euclidean space. To overcome the limitation on the relative velocity of observers, one should simply divide any set of observers into subsets of observers with non-relativistic relative velocities.\cite{partition} This means that the relativity theory can in no way avoid dealing with a variety of moving physical spaces or inertial reference frames.

The realization of Euclidean geometry with an aid of rigid bodies also fails over large scales due to the action of gravity. Then the use of Newtonian mechanics for the extension of Euclidean geometry
is again possible, especially since motions of two gravitating masses are similar to those of two opposite electric charges. Further extension of this scheme to include relativistic motions is beyond the scope of this article.

\subsection{Inadequacy of coordinate transformation between two frames}\label{coordinate transformation}
One of the consequences of adopting the idea of spacetime, originated from Ref.~\onlinecite{Minkowski1908} and generalized in Part II of  Ref.~\onlinecite{Einstein&Grosmann1913}, is the exclusive use of coordinates in describing physical relationships.

In particular, Minkowski space does make it redundant to provide the transformation of the base vectors between two inertial reference frames in addition to the transformation of the time and Cartesian coordinates of a spacetime point, associated with a physical event.\cite[\S 2.9]{Misner1973} Considering Cartesian coordinates in each inertial reference frame along with its associated time variable as fundamental quantities for each spatial space, a researcher has to either accept the existence of Minkowski space as another hypothesis or try to use the coordinate transformation
\begin{equation}\label{general transformation} 
\vec{\rho}^{(\mathrm{Bb})}=\mathbb{M}^{\mathrm{Bb}}_{\mathrm{Aa}}\vec{\rho}^{(\mathrm{Aa})}
\end{equation}
as a premise for the proof of the existence of Minkowski space.

In Eq.~\eqref{general transformation} the column vector
\[
\vec{\rho}^{(\mathrm{Ff})}=
\begin{pmatrix}
ct^{(\mathrm{Ff})}\\
x^{(\mathrm{Ff})}\\
y^{(\mathrm{Ff})}\\
z^{(\mathrm{Ff})}
\end{pmatrix}
\]
is made of the time and Cartesian coordinates of a physical event in an inertial reference frame f introduced in a physical space F, the matrix
\begin{equation}\label{generalized Lorentz matrix}
\mathbb{M}^{\mathrm{Bb}}_{\mathrm{Aa}}=
\mathbb{R}^{-1}\left(\vec{n}^{(\mathrm{b})}_{\mathrm{A}}\right)\mathbb{L}(v_{BA})\mathbb{R}\left(\vec{n}^{(\mathrm{a})}_{\mathrm{B}}\right)
\end{equation}
where
\[
\mathbb{L}(v)=
\begin{pmatrix}
\gamma(v) & -\gamma(v)v/c & 0 & 0\\
-\gamma(v)v/c & \gamma(v) & 0 & 0\\
0 & 0 & 1 & 0\\
0 & 0 & 0 & 1
\end{pmatrix}
\]
is the matrix of the historically original Lorentz transformation, now referred to as the special Lorentz transformation\cite[p.~41]{Moeller1955}, the matrix
\[
\mathbb{R}(\vec{n})=
\begin{pmatrix}
1&0&0&0\\
0&\sin\theta\cos\phi&\sin\theta\sin\phi&\cos\theta\\
0 &-\sin\phi&\cos\phi&0\\
0 &-\cos\theta\cos\phi&-\cos\theta\sin\phi&\sin\theta
\end{pmatrix}
\]
describes such rotation that \(\vec{e}_x=\mathbb{R}(\vec{n})\vec{n}\) for
\[
\vec{e}_x=\begin{pmatrix}
0\\
1\\
0\\
0 
\end{pmatrix},\quad
\vec{n}=\begin{pmatrix}
0\\
\sin\theta\cos\phi\\
\sin\theta\sin\phi\\
\cos\theta
\end{pmatrix},\quad 0\le\phi<2\pi, 0\le\theta<\pi.
\]

The column vector \(\vec{n}^{(\mathrm{f})}_{\mathrm{G}}\) in Eq.~\eqref{generalized Lorentz matrix} describes the direction of the velocity \(\vec{v}^{(\mathrm{Ff})}_{\mathrm{G}}=\vec{n}^{(\mathrm{f})}_{\mathrm{G}}v_{FG}\) of an G space in an inertial reference frame f, stationary in a F space.

The physically evident Eq.~\eqref{generalized Lorentz matrix} corresponds to the decomposition\cite{decomposition} of any physically reasonable transformation between two inertial reference frames into the product \(\mathbb{R}_1\mathbb{B}(\vec{v}_1)\) (or \(\mathbb{B}(\vec{v}_2) \mathbb{R}_2\)) where \(\mathbb{R}_1\) (or \(\mathbb{R}_2\)) is the matrix of a rotation while \(\mathbb{B}(\vec{v})\) is the matrix of the so called pure boost\cite{boost}, which itself can be decomposed\cite{boost decomposition} as \(\mathbb{B}(\vec{v})=\mathbb{R}^{-1}(\vec{n})\mathbb{L}(v)\mathbb{R}(\vec{n})\) for the velocity \(\vec{v}=v\vec{n}\).

The limiting case A=B=F, a=b=f entails that \(v_{FF}=0\) and
\begin{equation}\label{general reflexivity} 
\mathbb{M}^{\mathrm{Ff}}_{\mathrm{Ff}}=\mathbb{I}.
\end{equation}
The symmetry of exchanging two spaces (along with the frames they contain) yields \(v_{FG}=v_{GF}\) and
\begin{equation}\label{general symmetry} 
\mathbb{M}^{\mathrm{Ff}}_{\mathrm{Gg}}=\mathbb{M}^{\mathrm{Gg}}_{\mathrm{Ff}}.
\end{equation}

As soon as one establishes that for every \(\vec{v}^{(\mathrm{Aa})}_{\mathrm{G}}\) and \(\vec{v}^{(\mathrm{Gg})}_{\mathrm{B}}\) there is \(\vec{v}^{(\mathrm{Aa})}_{\mathrm{B}}\) such that
\begin{equation}\label{general transitivity}
\mathbb{M}^{\mathrm{Bb}}_{\mathrm{Aa}}=\mathbb{M}^{\mathrm{Bb}}_{\mathrm{Gg}} \mathbb{M}^{\mathrm{Gg}}_{\mathrm{Aa}},
\end{equation}
one can conclude that Eq.~\eqref{general transformation} is an equivalence relation since Eqs.~\eqref{general reflexivity}-\eqref{general transitivity} represent reflexivity, symmetry and transitivity of Eq.~\eqref{general transformation}. On denoting a corresponding equivalence class by \(\boldsymbol{\rho}\), one can finally get Minkowski space as a set of all possible \(\boldsymbol{\rho}\).

Apparently, the idea that \(\mathbb{M}^{(\mathrm{Ff})}_{(\mathrm{Gg})}\) represents a 4D pseudo-rotation provides little help in finding \(\vec{v}^{(\mathrm{Aa})}_{\mathrm{B}}\) for the given \(\vec{v}^{(\mathrm{Aa})}_{\mathrm{G}}\) and \(\vec{v}^{(\mathrm{Gg})}_{\mathrm{B}}\) in order to arrive at Eq.~\eqref{general transitivity} without addressing vectors in Minkowski space prematurely (i.e. without being circular.)

The limiting case A=B=G is more suggestive since it turns \(\mathbb{M}^{\mathrm{Bb}}_{\mathrm{Aa}}\) into the rotation in the 3D physical space G such that \(\vec{n}^{(\mathrm{b})}_{\mathrm{G}}=\mathbb{M}^{\mathrm{Gb}}_{\mathrm{Ga}}\vec{n}^{(\mathrm{a})}_{\mathrm{G}}\), and the corresponding equivalence class turns out to be a point of G, which is customarily marked by the Euclidean position vector \(\mathbf{r}\). However, the existence of the vector \(\mathbf{r}\) needs no reasoning based on rotations presented as transformations of Cartesian coordinates because Euclidean points and true spatial vectors have their own foundation in physics, as Section~\ref{physical spaces} indicates.

Since in the above evidently true limiting case Eq.~\eqref{general transitivity} yields some relationships between the components of \(\vec{n}^{\mathrm{f}}_{\mathrm{G}}\) in different frames, one can  hope that these relationships in conjunction of some ponderous decompositions of the matrices \(\mathbb{M}^{(\mathrm{Ff})}_{(\mathrm{Gg})}\) enable one to succeed in establishing Eq.~\eqref{general transitivity} in a general case. However, in order to get the desired decompositions, one has no way but to abandon the premature use of reference frames and turn to the transformation of the time \(t\) and the position vector \(\mathbf{r}\) of an event between physical spaces.

\subsection{Transformation between two physical spaces}\label{transformation}
Collisions between the particles that can interact over a short range only as well as their interceptions with features of the propagating electromagnetic field, such as rays and plane phase fronts, form a class of events that underlie the special relativity theory. As soon as one accepts that the events can be marked with Einsteinian time variable, explained in Section~\ref{Einsteinian time}, the principle of relativity along with the constancy of the speed of light lead one to the basic manifestations of the special relativity theory: length contraction, time dilation, time retardation, spatial transversal invariance.\cite[Sec.~III]{Wagner2016a} The coordinate-free description of these effects results in the transformation between physical spaces in the form of relationships between the times and physically essential components of the position vectors of a given event.

Let a superscript (F) of a quantity denote that the quantity is defined in a space F. In addition to Einsteinian time variable \(t^{(\mathrm{F})}\) and the position vector \(\mathbf{r}^{(\mathrm{F})}\) of a physical event in a space F, one can use the velocity \(\mathbf{v}^{(\mathrm{F})}_{\mathrm{G}}\) of another space G there.

Then the transformation of the time and the position vector of a physical event between the spaces A and B can be written as\cite[Sec.~IV]{Wagner2016a}
\begin{equation}\label{time relation AB}
t^{(\mathrm{B})}=\gamma_{ \mathrm{AB}}\left[t^{(\mathrm{A})}-\frac{(\mathbf{v}^{(\mathrm{A})}_{\mathrm{B}}\cdot\mathbf{r}^{(\mathrm{A})})}{c^2}\right],
\end{equation}
\begin{equation}\label{longitudinal relation AB} 
-\frac{ \left(\mathbf{r}^{(\mathrm{B})}\cdot\mathbf{v}^{(\mathrm{B})}_{\mathrm{A}}\right)}{v_{\mathrm{AB}} }=
\gamma _{\mathrm{AB}}\left[\frac{\left(\mathbf{r}^{(\mathrm{A})}\cdot\mathbf{v}^{(\mathrm{A})}_{\mathrm{B}}\right)}{v_{ \mathrm{AB}} }-v_{ \mathrm{AB}}t^{(\mathrm{A})}\right],
\end{equation}
\begin{equation}\label{transversal relation AB} 
\mathbf{r}^{(\mathrm{B})}-\frac{\left(\mathbf{r}^{(\mathrm{B})}\cdot\mathbf{v}^{(\mathrm{B})}_{\mathrm{A}}\right)\mathbf{v}^{(\mathrm{B})}_{\mathrm{A}}}{v_{\mathrm{AB}}^2}
\backsim
\mathbf{r}^{(\mathrm{A})}-\frac{\left(\mathbf{r}^{(\mathrm{A})}\cdot\mathbf{v}^{(\mathrm{A})}_{\mathrm{B}}\right)\mathbf{v}^{(\mathrm{A})}_{\mathrm{B}}}{v_{\mathrm{AB}}^2}.
\end{equation}
Since due to the symmetry of exchanging two spaces \(\left|\mathbf{v}^{(\mathrm{A})}_{\mathrm{B}}\right|=\left|\mathbf{v}^{(\mathrm{B})}_{\mathrm{A}}\right|\), the additional notation
\begin{equation}\label{velocity symmetry AB} 
v_{\mathrm{AB}}
\equiv\left|\mathbf{v}^{(\mathrm{A})}_{\mathrm{B}}\right|=\left|\mathbf{v}^{(\mathrm{B})}_{\mathrm{A}}\right|,\quad\gamma_{\mathrm{AB}}\equiv\gamma(v_{\mathrm{AB}})
\end{equation}
with the aid of the standard function
\begin{equation*}
\gamma(v)\equiv\frac{1}{\sqrt{1-v^2/c^2}}
\end{equation*}
is used here and hereinafter.

In contrast to the relation ``\(=\)", the relation ``\(\backsim\)" connects quantities in different spaces. Still, the principle of relativity implies that\cite[Sec.~III.B]{Wagner2016a}
\begin{equation}\label{addition property}
\mathbf{f}^{(\mathrm{A})}_1\backsim\mathbf{g}^{(\mathrm{B})}_1\text{ and }\mathbf{f}^{(\mathrm{A})}_2\backsim\mathbf{g}^{(\mathrm{B})}_2\text{ entail } \mathbf{f}^{(\mathrm{A})}_1+\mathbf{f}^{(\mathrm{A})}_2\backsim\mathbf{g}^{(\mathrm{B})}_1+\mathbf{g}^{(\mathrm{B})}_2
\end{equation}
and
\begin{equation}\label{dot property}
\mathbf{f}^{(\mathrm{A})}_3\backsim\mathbf{g}^{(\mathrm{B})}_3\text{ and }\mathbf{f}^{(\mathrm{A})}_4\backsim\mathbf{g}^{(\mathrm{B})}_4\text{ entail } \left(\mathbf{f}^{(\mathrm{A})}_3\cdot\mathbf{f}^{(\mathrm{A})}_4\right)= \left(\mathbf{g}^{(\mathrm{B})}_3\cdot\mathbf{g}^{(\mathrm{B})}_4\right)
\end{equation}
for any spatial vector \(\mathbf{f}^{(\mathrm{A})}_i\) in the space A and its counterpart \(\mathbf{g}^{(\mathrm{B})}_i\) in the space B.

It is worth remarking that the transformation rules \eqref{longitudinal relation AB} and \eqref{transversal relation AB} differ from the vector-like relationship presented in the literature\cite{vector-like transformation} because the latter actually deals with column vectors made of Cartesian coordinates of true vectors and appears identical to the so called boost coordinate transformation.\cite{boost transformation} However, unlike Eq.~\eqref{generalized Lorentz matrix}, this transformation implies the special, apparently unphysical, choice of the coordinate systems where the column vectors \(-\vec{v}^{(\mathrm{A})}_{\mathrm{B}}\) and \(\vec{v}^{(\mathrm{B})}_{\mathrm{A}}\) are equal. Therefore, such a transformation proves to be completely irrelevant to the question under consideration.

In a more compact form, one can write the transformation \eqref{time relation AB}-\eqref{transversal relation AB} as the mapping
\begin{equation}\label{tr relation}
\begin{pmatrix}
ct^{(\mathrm{B})}\\
\mathbf{r}^{(\mathrm{B})}
\end{pmatrix}
\leftrightarrow
\mathbf{M}^{(\mathrm{B})}_{(\mathrm{A})}\odot\begin{pmatrix}
ct^{(\mathrm{A})}\\
\mathbf{r}^{(\mathrm{A})}
\end{pmatrix}
\end{equation}
where
\begin{equation}\label{vector transformation matrix}
\mathbf{M}^{(\mathrm{B})}_{(\mathrm{A})}\equiv\begin{pmatrix}
\gamma_{\mathrm{AB}} &-\gamma_{\mathrm{AB}}\mathbf{v}^{(\mathrm{A})}_{\mathrm{B}}/c\\
\gamma_{\mathrm{AB}}\mathbf{v}^{(\mathrm{B})}_{\mathrm{A}}/c&1-\gamma_{\mathrm{AB}} \mathbf{v}^{(\mathrm{B})}_{\mathrm{A}}\otimes\mathbf{v}^{(\mathrm{A})}_{\mathrm{B}}/v_{\mathrm{AB}}^2
\end{pmatrix},
\end{equation}
the symbol \(\leftrightarrow\) unites the meaning of \(=\) and the meaning of \(\backsim\), the symbol \(\odot\) unites the meaning of the usual product of two numbers and the meaning of the dot product of two spatial vectors,  the symbol \(\otimes\) denotes the dyadic (outer) product.

As soon as one shows that Eq.~\eqref{tr relation} is an equivalence relation, one arrives at a spacetime point \(\boldsymbol{\rho}\) as an equivalence class of columns
\(\begin{pmatrix}
ct\\
\mathbf{r}
\end{pmatrix}\)
that indicate to that spacetime point in their spaces. Then the set of all \(\boldsymbol{\rho}\) makes Minkowski space.

\section{Proof of existence of Minkowski space}\label{proof}
\subsection{What requires a verification}\label{transitivity}
The mapping \eqref{tr relation} is evidently reflexive. To show its symmetry, one could resolve Eqs. \eqref{time relation AB} and \eqref{longitudinal relation AB} with respect to \(t^{(\mathrm{A})}\) and \((\mathbf{v}^{(\mathrm{A})}_{\mathrm{B}}\cdot\mathbf{r}^{(\mathrm{A})})\), taking Eq.~\eqref{velocity symmetry AB} into account. However, this action would be redundant since the symmetry of exchanging two spaces is physically evident and should be considered as a premise rather than an inference.
Thus, to show that the mapping \eqref{tr relation} is an equivalence relation, one need verifying only its transitivity.

Although Eq.~\eqref{general transitivity} may tempt someone to present the transitivity of \eqref{tr relation} as
\begin{equation}\label{general vectorial transitivity} 
\mathbf{M}^{(\mathrm{B})}_{(\mathrm{A})}=\mathbf{M}^{(\mathrm{B})}_{(\mathrm{G})}\odot \mathbf{M}^{(\mathrm{G})}_{(\mathrm{A})},
\end{equation}
it actually means that
\begin{equation}\label{time relation BG} 
t^{(\mathrm{B})}=\gamma_{\mathrm{BG}}\left[t^{(\mathrm{G})}-\frac{\left(\mathbf{v}^{(\mathrm{G})}_{\mathrm{B}}\cdot\mathbf{r}^{(\mathrm{G})}\right)}{c^2}\right],
\end{equation}
\begin{equation}\label{longitudinal relation BG} 
-\frac{ \left(\mathbf{v}^{(\mathrm{B})}_{\mathrm{G}}\cdot\mathbf{r}^{(\mathrm{B})}\right) }{ v_{\mathrm{BG}} }=
\gamma _{\mathrm{BG}}\left[\frac{\left(\mathbf{v}^{(\mathrm{G})}_{\mathrm{B}}\cdot\mathbf{r}^{(\mathrm{G})}\right)}{v_{ \mathrm{BG}} }-v_{ \mathrm{BG}}t^{(\mathrm{G})}\right],
\end{equation}
\begin{equation}\label{transversal relation BG} 
\mathbf{r}^{(\mathrm{B})}-\frac{\left(\mathbf{r}^{(\mathrm{B})}\cdot\mathbf{v}^{(\mathrm{B})}_{\mathrm{G}}\right)\mathbf{v}^{(\mathrm{B})}_{\mathrm{G}}}{v_{\mathrm{BG}}^2}
\backsim
\mathbf{r}^{(\mathrm{G})}-\frac{\left(\mathbf{r}^{(\mathrm{G})}\cdot\mathbf{v}^{(\mathrm{G})}_{\mathrm{B}}\right)\mathbf{v}^{(\mathrm{G})}_{\mathrm{B}}}{v_{\mathrm{BG}}^2}
\end{equation}
and
\begin{equation}\label{time relation AG}
t^{(\mathrm{G})}=\gamma_{ \mathrm{AG}}\left[t^{(\mathrm{A})}-\frac{\left(\mathbf{v}^{(\mathrm{A})}_{\mathrm{G}}\cdot\mathbf{r}^{(\mathrm{A})}\right)}{c^2}\right],
\end{equation}

\begin{equation}\label{longitudinal relation AG}
-\frac{ \left(\mathbf{v}^{(\mathrm{G})}_{\mathrm{A}}\cdot\mathbf{r}^{(\mathrm{G})}\right) }{ v_{\mathrm{AG}} }=
\gamma _{\mathrm{AG}}\left[\frac{\left(\mathbf{v}^{(\mathrm{A})}_{\mathrm{G}}\cdot\mathbf{r}^{(\mathrm{A})}\right)}{v_{ \mathrm{AG}} }-v_{ \mathrm{AG}}t^{(\mathrm{A})}\right],
\end{equation}
\begin{equation}\label{transversal relation AG} 
\mathbf{r}^{(\mathrm{G})}-\frac{\left(\mathbf{r}^{(\mathrm{G})}\cdot\mathbf{v}^{(\mathrm{G})}_{\mathrm{A}}\right)\mathbf{v}^{(\mathrm{G})}_{\mathrm{A}}}{v_{\mathrm{AG}}^2}
\backsim
\mathbf{r}^{(\mathrm{A})}-\frac{\left(\mathbf{r}^{(\mathrm{A})}\cdot\mathbf{v}^{(\mathrm{A})}_{\mathrm{G}}\right)\mathbf{v}^{(\mathrm{A})}_{\mathrm{G}}}{v_{\mathrm{AG}}^2}
\end{equation}
entail Eqs.~\eqref{time relation AB}-\eqref{transversal relation AB}.

\subsection{Auxiliary relationships for the velocities of spaces}\label{auxiliary velocities}
In accordance with the definition of the velocities \(\mathbf{v}^{(\mathrm{A})}_{\mathrm{G}}\) and \(\mathbf{v}^{(\mathrm{B})}_{\mathrm{G}}\), the motion \(\mathbf{r}^{(\mathrm{G})}=0\) of a reference point in the space G is observed as the motion \(\mathbf{r}^{(\mathrm{A})}=\mathbf{v}^{(\mathrm{A})}_{\mathrm{G}}t^{(\mathrm{A})}\) in the space A and as the motion \(\mathbf{r}^{(\mathrm{B})}=\mathbf{v}^{(\mathrm{B})}_{\mathrm{G}}t^{(\mathrm{B})}\) in the space B.
With applying the transformation \eqref{time relation AB}-\eqref{transversal relation AB}
to this set of events, one finds
\begin{equation}\label{G motion time AB}
t^{(\mathrm{B})}=\gamma_{ \mathrm{AB}}\left[1-\frac{\left(\mathbf{v}^{(\mathrm{A})}_{\mathrm{B}}\cdot\mathbf{v}^{(\mathrm{A})}_{\mathrm{G}}\right)}{c^2}\right]t^{(\mathrm{A})},
\end{equation}
\begin{equation}\label{G motion along AB}
-\frac{ \left(\mathbf{v}^{(\mathrm{B})}_{\mathrm{A}}\cdot\mathbf{v}^{(\mathrm{B})}_{\mathrm{G}}\right) }{ v_{\mathrm{AB}} }t^{(\mathrm{B})}=
\gamma _{\mathrm{AB}}\left[\frac{\left(\mathbf{v}^{(\mathrm{A})}_{\mathrm{B}}\cdot\mathbf{v}^{(\mathrm{A})}_{\mathrm{G}}\right)}{v_{ \mathrm{AB}} }-v_{ \mathrm{AB}}\right]t^{(\mathrm{A})},
\end{equation}
\begin{equation}\label{G motion across AB}
\left[\mathbf{v}^{(\mathrm{B})}_{\mathrm{G}}-\frac{\left(\mathbf{v}^{(\mathrm{B})}_{\mathrm{G}}\cdot\mathbf{v}^{(\mathrm{B})}_{\mathrm{A}}\right)\mathbf{v}^{(\mathrm{B})}_{\mathrm{A}}}{v_{\mathrm{AB}}^2}\right]t^{(\mathrm{B})}
\backsim
\left[\mathbf{v}^{(\mathrm{A})}_{\mathrm{G}}-\frac{\left(\mathbf{v}^{(\mathrm{A})}_{\mathrm{G}}\cdot\mathbf{v}^{(\mathrm{A})}_{\mathrm{B}}\right)\mathbf{v}^{(\mathrm{A})}_{\mathrm{B}}}{v_{\mathrm{AB}}^2}\right]t^{(\mathrm{A})}.
\end{equation}

Transposing A\(\leftrightarrow\)B in Eq.~\eqref{G motion time AB} (which means the use of the transformation inverse to \eqref{time relation AB}-\eqref{transversal relation AB} in the above calculation) yields
\begin{equation}\label{G motion time BA}
t^{(\mathrm{A})}=\gamma_{\mathrm{AB}}\left[1-\frac{\left(\mathbf{v}^{(\mathrm{B})}_{\mathrm{A}}\cdot\mathbf{v}^{(\mathrm{B})}_{\mathrm{G}}\right)}{c^2}\right]t^{(\mathrm{B})}.
\end{equation}
Then one can eliminate \(t^{(\mathrm{A})}\) and \(t^{(\mathrm{B})}\) in Eq.~\eqref{G motion time AB} and Eq.~\eqref{G motion time BA} to obtain
\begin{equation}\label{gamma AB velocities}
1=\gamma_{\mathrm{AB}}^2\left[1-\frac{\left(\mathbf{v}^{(\mathrm{A})}_{\mathrm{B}}\cdot\mathbf{v}^{(\mathrm{A})}_{\mathrm{G}}\right)}{c^2}\right]\left[1-\frac{\left(\mathbf{v}^{(\mathrm{B})}_{\mathrm{A}}\cdot\mathbf{v}^{(\mathrm{B})}_{\mathrm{G}}\right)}{c^2}\right].
\end{equation}

With transposing B\(\leftrightarrow\)G and A\(\leftrightarrow\)G in Eq.~\eqref{gamma AB velocities}, one can also find
\begin{equation}\label{gamma AG velocities}
1=\gamma_{ \mathrm{AG}}^2\left[1-\frac{\left(\mathbf{v}^{(\mathrm{A})}_{\mathrm{B}}\cdot\mathbf{v}^{(\mathrm{A})}_{\mathrm{G}}\right)}{c^2}\right]\left[1-\frac{\left(\mathbf{v}^{(\mathrm{G})}_{\mathrm{B}}\cdot\mathbf{v}^{(\mathrm{G})}_{\mathrm{A}}\right)}{c^2}\right]
\end{equation}
and
\begin{equation}\label{gamma BG velocities}
1=\gamma_{ \mathrm{BG}}^2\left[1-\frac{\left(\mathbf{v}^{(\mathrm{B})}_{\mathrm{A}}\cdot\mathbf{v}^{(\mathrm{B})}_{\mathrm{G}}\right)}{c^2}\right]\left[1-\frac{\left(\mathbf{v}^{(\mathrm{G})}_{\mathrm{B}}\cdot\mathbf{v}^{(\mathrm{G})}_{\mathrm{A}}\right)}{c^2}\right].
\end{equation}

In a similar manner,  one can start from Eq.~\eqref{G motion time AB} and Eq.~\eqref{G motion along AB} to go to the relationships
\begin{equation}\label{velocities product A-G} 
\frac{\left(\mathbf{v}^{(\mathrm{A})}_{\mathrm{B}}\cdot\mathbf{v}^{(\mathrm{A})}_{\mathrm{G}}\right)}{v_{ \mathrm{AG}}^2}\left[1-\frac{\left(\mathbf{v}^{(\mathrm{G})}_{\mathrm{B}}\cdot\mathbf{v}^{(\mathrm{G})}_{\mathrm{A}}\right)}{c^2}\right] =
1-\frac{\left(\mathbf{v}^{(\mathrm{G})}_{\mathrm{B}}\cdot\mathbf{v}^{(\mathrm{G})}_{\mathrm{A}}\right)}{v_{ \mathrm{AG}}^2},
\end{equation}

\begin{equation}\label{velocities product A-B}
\frac{\left(\mathbf{v}^{(\mathrm{A})}_{\mathrm{B}}\cdot\mathbf{v}^{(\mathrm{A})}_{\mathrm{G}}\right)}{v_{ \mathrm{AB}}^2}\left[1-\frac{\left(\mathbf{v}^{(\mathrm{B})}_{\mathrm{G}}\cdot\mathbf{v}^{(\mathrm{B})}_{\mathrm{A}}\right)}{c^2}\right]
= 1-\frac{\left(\mathbf{v}^{(\mathrm{B})}_{\mathrm{G}}\cdot\mathbf{v}^{(\mathrm{B})}_{\mathrm{A}}\right)}{v_{ \mathrm{AB}}^2},
\end{equation}

\begin{equation}\label{velocities product B-G}
\frac{\left(\mathbf{v}^{(\mathrm{B})}_{\mathrm{A}}\cdot\mathbf{v}^{(\mathrm{B})}_{\mathrm{G}}\right)}{v_{ \mathrm{BG}}^2}\left[1-\frac{\left(\mathbf{v}^{(\mathrm{G})}_{\mathrm{B}}\cdot\mathbf{v}^{(\mathrm{G})}_{\mathrm{A}}\right)}{c^2}\right]
= 1-\frac{\left(\mathbf{v}^{(\mathrm{G})}_{\mathrm{B}}\cdot\mathbf{v}^{(\mathrm{G})}_{\mathrm{A}}\right)}{v_{ \mathrm{BG}}^2}.
\end{equation}

It is easy see that
 Eq.~\eqref{G motion time AB} and Eq.~\eqref{G motion across AB} entail
\begin{equation}\label{velocity transversal relation AGB}
\mathbf{v}^{(\mathrm{B})}_{\mathrm{G}}-\frac{\left(\mathbf{v}^{(\mathrm{B})}_{\mathrm{G}}\cdot\mathbf{v}^{(\mathrm{B})}_{\mathrm{A}}\right)\mathbf{v}^{(\mathrm{B})}_{\mathrm{A}}}{v_{\mathrm{AB}}^2}
\backsim
\frac{\mathbf{v}^{(\mathrm{A})}_{\mathrm{G}}-\frac{\left(\mathbf{v}^{(\mathrm{A})}_{\mathrm{G}}\cdot\mathbf{v}^{(\mathrm{A})}_{\mathrm{B}}\right)\mathbf{v}^{(\mathrm{A})}_{\mathrm{B}}}{v_{\mathrm{AB}}^2}}{\gamma_{ \mathrm{AB}}\left[1-\frac{\left(\mathbf{v}^{(\mathrm{A})}_{\mathrm{B}}\cdot\mathbf{v}^{(\mathrm{A})}_{\mathrm{G}}\right)}{c^2}\right]}.
\end{equation}
In addition, transposing B\(\leftrightarrow\)G turns Eq.~\eqref{velocity transversal relation AGB} into
\begin{equation}\label{velocity transversal relation ABG}
\mathbf{v}^{(\mathrm{G})}_{\mathrm{B}}-\frac{\left(\mathbf{v}^{(\mathrm{G})}_{\mathrm{B}}\cdot\mathbf{v}^{(\mathrm{G})}_{\mathrm{A}}\right)\mathbf{v}^{(\mathrm{G})}_{\mathrm{A}}}{v_{\mathrm{AG}}^2}
\backsim
\frac{\mathbf{v}^{(\mathrm{A})}_{\mathrm{B}}-\frac{\left(\mathbf{v}^{(\mathrm{A})}_{\mathrm{B}}\cdot\mathbf{v}^{(\mathrm{A})}_{\mathrm{G}}\right)\mathbf{v}^{(\mathrm{A})}_{\mathrm{G}}}{v_{\mathrm{AG}}^2}}{\gamma_{ \mathrm{AG}}\left[1-\frac{\left(\mathbf{v}^{(\mathrm{A})}_{\mathrm{B}}\cdot\mathbf{v}^{(\mathrm{A})}_{\mathrm{G}}\right)}{c^2}\right]}.
\end{equation}

Combining Eqs.~\eqref {gamma AB velocities}-\eqref{gamma BG velocities}
yields 
\begin{equation}\label{gamma AB}
\gamma_{\mathrm{AB}}=\gamma_{\mathrm{AG}}\gamma_{ \mathrm{BG}}\left[1-\frac{\left(\mathbf{v}^{(\mathrm{G})}_{\mathrm{B}}\cdot\mathbf{v}^{(\mathrm{G})}_{\mathrm{A}}\right)}{c^2}\right].
\end{equation}

There are the identities possible due to the symmetry of exchanging the spaces:
Since transposing any two of A and B and G does not change the r.h.s. of the equation
\[
\gamma_{\mathrm{AG}}^{-2}\left[1-\frac{\left(\mathbf{v}^{(\mathrm{B})}_{\mathrm{G}}\cdot\mathbf{v}^{(\mathrm{B})}_{\mathrm{A}}\right)}{c^2}\right]=\frac{1}{\gamma_{\mathrm{AB}}\gamma_{\mathrm{AG}}\gamma _{\mathrm{BG}}},
\]
it entails the identities
\begin{equation}\label{transposition symmetry BB-GG}
\gamma_{\mathrm{AG}}^{-2}\left[1-\frac{\left(\mathbf{v}^{(\mathrm{B})}_{\mathrm{G}}\cdot\mathbf{v}^{(\mathrm{B})}_{\mathrm{A}}\right)}{c^2}\right]=\gamma_{\mathrm{AB}}^{-2}\left[1-\frac{\left(\mathbf{v}^{(\mathrm{G})}_{\mathrm{B}}\cdot\mathbf{v}^{(\mathrm{G})}_{\mathrm{A}}\right)}{c^2}\right]
\end{equation}
and
\begin{equation}\label{transposition symmetry BB-AA}
\gamma_{\mathrm{AG}}^{-2}\left[1-\frac{\left(\mathbf{v}^{(\mathrm{B})}_{\mathrm{G}}\cdot\mathbf{v}^{(\mathrm{B})}_{\mathrm{A}}\right)}{c^2}\right]=\gamma_{\mathrm{GB}}^{-2}\left[1-\frac{\left(\mathbf{v}^{(\mathrm{A})}_{\mathrm{B}}\cdot\mathbf{v}^{(\mathrm{A})}_{\mathrm{G}}\right)}{c^2}\right].
\end{equation}

Similarly, the equation
\[
\frac{v_{\mathrm{AG}}^2v_{\mathrm{BG}}^2-\left(\mathbf{v}^{(\mathrm{G})}_{\mathrm{B}}\cdot\mathbf{v}^{(\mathrm{G})}_{\mathrm{A}}\right)^2
}{\gamma_{\mathrm{AB}}^2c^4}=\frac{2}{\gamma_{\mathrm{AB}}\gamma_{\mathrm{AG}}\gamma _{\mathrm{BG}}}-\left(
\frac{1}{\gamma_{\mathrm{AB}}^2\gamma_{\mathrm{AG}}^2} +\frac{1}{\gamma_{\mathrm{AB}}^2\gamma _{\mathrm{BG}}^2} +\frac{1}{\gamma_{\mathrm{AG}}^2\gamma _{\mathrm{BG}}^2}
\right)+\frac{1}{\gamma_{\mathrm{AB}}^2\gamma_{\mathrm{AG}}^2\gamma _{\mathrm{BG}}^2}
\]
produces
\begin{equation}\label{transposition symmetry GG2-BB2}
\frac{v_{\mathrm{AG}}^2v_{\mathrm{BG}}^2-\left(\mathbf{v}^{(\mathrm{G})}_{\mathrm{B}}\cdot\mathbf{v}^{(\mathrm{G})}_{\mathrm{A}}\right)^2
}{\gamma_{\mathrm{AB}}^2} = \frac{v_{\mathrm{AB}}^2v_{\mathrm{BG}}^2-\left(\mathbf{v}^{(\mathrm{B})}_{\mathrm{G}}\cdot\mathbf{v}^{(\mathrm{B})}_{\mathrm{A}}\right)^2
}{\gamma_{\mathrm{AG}}^2}
\end{equation}
and
\begin{equation}\label{transposition symmetry BB2-AA2}
\frac{v_{\mathrm{AB}}^2v_{\mathrm{BG}}^2-\left(\mathbf{v}^{(\mathrm{B})}_{\mathrm{G}}\cdot\mathbf{v}^{(\mathrm{B})}_{\mathrm{A}}\right)^2
}{\gamma_{\mathrm{AG}}^2}
= \frac{v_{\mathrm{AB}}^2v_{\mathrm{AG}}^2-\left(\mathbf{v}^{(\mathrm{A})}_{\mathrm{G}}\cdot\mathbf{v}^{(\mathrm{A})}_{\mathrm{B}}\right)^2
}{\gamma_{\mathrm{BG}}^2}.
\end{equation}

\subsection{Auxiliary expressions for a longitudinal length}\label{auxiliary length}
The transformation rule \eqref{longitudinal relation BG} presents a length \(\left(\mathbf{v}^{(\mathrm{G})}_{\mathrm{B}}\cdot\mathbf{r}^{(\mathrm{G})}\right)/v_{\mathrm{BG}}\) along the direction of \(\mathbf{v}^{(\mathrm{G})}_{\mathrm{B}}\) in terms of quantities defined in a space G, associated with that direction.
However, the calculations in the following sections require to express a length along a given boost direction via quantities defined in an arbitrary third space.
In other words, one needs to express the dot product \(\left(\mathbf{v}^{(\mathrm{G})}_{\mathrm{B}}\cdot\mathbf{r}^{(\mathrm{G})}\right)\) in the space A.

To do the required calculation, one cannot but address the decompositions
\[
\mathbf{r}^{(\mathrm{G})}=\mathbf{r}^{(\mathrm{G})}-\frac{\left(\mathbf{r}^{(\mathrm{G})}\cdot\mathbf{v}^{(\mathrm{G})}_{\mathrm{A}}\right)\mathbf{v}^{(\mathrm{G})}_{\mathrm{A}}}{v_{\mathrm{AG}}^2} + \frac{\left(\mathbf{r}^{(\mathrm{G})}\cdot\mathbf{v}^{(\mathrm{G})}_{\mathrm{A}}\right)\mathbf{v}^{(\mathrm{G})}_{\mathrm{A}}}{v_{\mathrm{AG}}^2}
\]
and
\[
\mathbf{v}^{(\mathrm{G})}_{\mathrm{B}}=\mathbf{v}^{(\mathrm{G})}_{\mathrm{B}}-\frac{\left(\mathbf{v}^{(\mathrm{G})}_{\mathrm{B}}\cdot\mathbf{v}^{(\mathrm{G})}_{\mathrm{A}}\right)\mathbf{v}^{(\mathrm{G})}_{\mathrm{A}}}{v_{\mathrm{AG}}^2} + \frac{\left(\mathbf{v}^{(\mathrm{G})}_{\mathrm{B}}\cdot\mathbf{v}^{(\mathrm{G})}_{\mathrm{A}}\right)\mathbf{v}^{(\mathrm{G})}_{\mathrm{A}}}{v_{\mathrm{AG}}^2}
\]
and then apply the transformation rule \eqref{longitudinal relation AG}, the relations \eqref{transversal relation AG} and \eqref{velocity transversal relation ABG} with an aid of \eqref{dot property}:

\[
\left(\mathbf{v}^{(\mathrm{G})}_{\mathrm{B}}\cdot\mathbf{r}^{(\mathrm{G})}\right)=
\]
\[
=\left(\left[\mathbf{v}^{(\mathrm{G})}_{\mathrm{B}}-\frac{\left(\mathbf{v}^{(\mathrm{G})}_{\mathrm{B}}\cdot\mathbf{v}^{(\mathrm{G})}_{\mathrm{A}}\right)\mathbf{v}^{(\mathrm{G})}_{\mathrm{A}}}{v_{\mathrm{AG}}^2} \right]\cdot\left[\mathbf{r}^{(\mathrm{G})}-\frac{\left(\mathbf{r}^{(\mathrm{G})}\cdot\mathbf{v}^{(\mathrm{G})}_{\mathrm{A}}\right)\mathbf{v}^{(\mathrm{G})}_{\mathrm{A}}}{v_{\mathrm{AG}}^2}\right]\right)+\frac{\left(\mathbf{v}^{(\mathrm{G})}_{\mathrm{B}}\cdot\mathbf{v}^{(\mathrm{G})}_{\mathrm{A}}\right) \left(\mathbf{v}^{(\mathrm{G})}_{\mathrm{A}} \cdot \mathbf{r}^{(\mathrm{G})} \right) }{v_{\mathrm{AG}}^2}
\]
\[
=\frac{\left(\mathbf{v}^{(\mathrm{A})}_{\mathrm{B}}\cdot\mathbf{r}^{(\mathrm{A})}\right)-\frac{\left(\mathbf{v}^{(\mathrm{A})}_{\mathrm{G}}\cdot\mathbf{v}^{(\mathrm{A})}_{\mathrm{B}}\right)}{v_{\mathrm{AG}}^2}\left(\mathbf{v}^{(\mathrm{A})}_{\mathrm{G}}\cdot\mathbf{r}^{(\mathrm{A})}\right)}{\gamma_{\mathrm{AG}}\left[1-\frac{\left(\mathbf{v}^{(\mathrm{A})}_{\mathrm{B}}\cdot\mathbf{v}^{(\mathrm{A})}_{\mathrm{G}}\right)}{c^2}\right]}-\frac{\left(\mathbf{v}^{(\mathrm{G})}_{\mathrm{B}}\cdot\mathbf{v}^{(\mathrm{G})}_{\mathrm{A}}\right)}{v_{\mathrm{AG}}}\gamma _{\mathrm{AG}}\left[\frac{\left(\mathbf{v}^{(\mathrm{A})}_{\mathrm{G}}\cdot\mathbf{r}^{(\mathrm{A})}\right)}{v_{\mathrm{AG}} }-v_{\mathrm{AG}}t^{(\mathrm{A})}\right]=
\]
\[
=\gamma _{\mathrm{AG}}\left(\mathbf{v}^{(\mathrm{G})}_{\mathrm{B}}\cdot\mathbf{v}^{(\mathrm{G})}_{\mathrm{A}}\right)t^{(\mathrm{A})}+\gamma_{ \mathrm{AG}}\left[1-\frac{\left(\mathbf{v}^{(\mathrm{G})}_{\mathrm{B}}\cdot\mathbf{v}^{(\mathrm{G})}_{\mathrm{A}}\right)}{c^2}\right]\left(\mathbf{v}^{(\mathrm{A})}_{\mathrm{B}}\cdot\mathbf{r}^{(\mathrm{A})}\right)-
\]
\[
-\gamma_{ \mathrm{AG}}\left[1-\frac{\left(\mathbf{v}^{(\mathrm{G})}_{\mathrm{B}}\cdot\mathbf{v}^{(\mathrm{G})}_{\mathrm{A}}\right)}{c^2}\right]\frac{\left(\mathbf{v}^{(\mathrm{A})}_{\mathrm{G}}\cdot\mathbf{v}^{(\mathrm{A})}_{\mathrm{B}}\right)}{v_{\mathrm{AG}}^2}\left(\mathbf{v}^{(\mathrm{A})}_{\mathrm{G}}\cdot\mathbf{r}^{(\mathrm{A})}\right)-\frac{\left(\mathbf{v}^{(\mathrm{G})}_{\mathrm{B}}\cdot\mathbf{v}^{(\mathrm{G})}_{\mathrm{A}}\right)\left(\mathbf{v}^{(\mathrm{A})}_{\mathrm{G}}\cdot\mathbf{r}^{(\mathrm{A})}\right)}{v_{\mathrm{AG}}^2}\gamma _{\mathrm{AG}}=
\]
\[
=\gamma _{\mathrm{AG}}\left(\mathbf{v}^{(\mathrm{G})}_{\mathrm{B}}\cdot\mathbf{v}^{(\mathrm{G})}_{\mathrm{A}}\right)t^{(\mathrm{A})}-\gamma_{ \mathrm{AG}}\left(\mathbf{v}^{(\mathrm{A})}_{\mathrm{G}}\cdot\mathbf{r}^{(\mathrm{A})}\right)+\gamma_{ \mathrm{AG}}\left[1-\frac{\left(\mathbf{v}^{(\mathrm{G})}_{\mathrm{B}}\cdot\mathbf{v}^{(\mathrm{G})}_{\mathrm{A}}\right)}{c^2}\right]\left(\mathbf{v}^{(\mathrm{A})}_{\mathrm{B}}\cdot\mathbf{r}^{(\mathrm{A})}\right).
\]
Here the identities \eqref{gamma AG velocities} and
\eqref{velocities product A-G} are also used.

Thus,
\[\left(\mathbf{v}^{(\mathrm{G})}_{\mathrm{B}}\cdot\mathbf{r}^{(\mathrm{G})}\right)=\]
\begin{equation}\label{rGvGB in A}
=\gamma _{\mathrm{AG}}\left(\mathbf{v}^{(\mathrm{G})}_{\mathrm{B}}\cdot\mathbf{v}^{(\mathrm{G})}_{\mathrm{A}}\right)t^{(\mathrm{A})}-\gamma_{ \mathrm{AG}}\left(\mathbf{v}^{(\mathrm{A})}_{\mathrm{G}}\cdot\mathbf{r}^{(\mathrm{A})}\right)+\gamma_{ \mathrm{AG}}\left[1-\frac{\left(\mathbf{v}^{(\mathrm{G})}_{\mathrm{B}}\cdot\mathbf{v}^{(\mathrm{G})}_{\mathrm{A}}\right)}{c^2}\right]\left(\mathbf{v}^{(\mathrm{A})}_{\mathrm{B}}\cdot\mathbf{r}^{(\mathrm{A})}\right).
\end{equation}
Transposing B\(\leftrightarrow\)A and then G\(\leftrightarrow\)B turns this equation into
\[\left(\mathbf{v}^{(\mathrm{B})}_{\mathrm{A}}\cdot\mathbf{r}^{(\mathrm{B})}\right)=\]
\begin{equation}\label{rBvBA in G}
=\gamma _{\mathrm{BG}}\left(\mathbf{v}^{(\mathrm{B})}_{\mathrm{A}}\cdot\mathbf{v}^{(\mathrm{B})}_{\mathrm{G}}\right)t^{(\mathrm{G})}-\gamma _{\mathrm{BG}} \left(\mathbf{v}^{(\mathrm{G})}_{\mathrm{B}}\cdot\mathbf{r}^{(\mathrm{G})}\right)+\gamma_{ \mathrm{BG}}\left[1-\frac{\left(\mathbf{v}^{(\mathrm{B})}_{\mathrm{G}}\cdot\mathbf{v}^{(\mathrm{B})}_{\mathrm{A}}\right)}{c^2}\right]\left(\mathbf{v}^{(\mathrm{G})}_{\mathrm{A}}\cdot\mathbf{r}^{(\mathrm{G})}\right).
\end{equation}

Due to the transformation rule \eqref{longitudinal relation AG} one has
\begin{equation}\label{rGvGA in A}
\left(\mathbf{v}^{(\mathrm{G})}_{\mathrm{A}} \cdot \mathbf{r}^{(\mathrm{G})} \right)=-\gamma _{\mathrm{AG}}\left[ \left(\mathbf{v}^{(\mathrm{A})}_{\mathrm{G}}\cdot\mathbf{r}^{(\mathrm{A})}\right)-v_{ \mathrm{AG}}^2t^{(\mathrm{A})}\right]
\end{equation}

With transposing A\(\leftrightarrow\)B, one can obtain
\begin{equation}\label{rGvGB in B}
\left(\mathbf{v}^{(\mathrm{G})}_{\mathrm{B}}\cdot\mathbf{r}^{(\mathrm{G})}\right)=-\gamma _{\mathrm{BG}}\left[ \left(\mathbf{v}^{(\mathrm{B})}_{\mathrm{G}}\cdot\mathbf{r}^{(\mathrm{B})}\right)-v_{ \mathrm{BG}}^2t^{(\mathrm{B})}\right]
\end{equation}
and
\[\left(\mathbf{v}^{(\mathrm{G})}_{\mathrm{A}}\cdot\mathbf{r}^{(\mathrm{G})}\right)=\]
\begin{equation}\label{rGvGA in B}
=\gamma _{\mathrm{BG}}\left(\mathbf{v}^{(\mathrm{G})}_{\mathrm{B}}\cdot\mathbf{v}^{(\mathrm{G})}_{\mathrm{A}}\right)t^{(\mathrm{B})}-\gamma_{ \mathrm{BG}}\left(\mathbf{v}^{(\mathrm{B})}_{\mathrm{G}}\cdot\mathbf{r}^{(\mathrm{B})}\right)+\gamma_{ \mathrm{BG}}\left[1-\frac{\left(\mathbf{v}^{(\mathrm{G})}_{\mathrm{B}}\cdot\mathbf{v}^{(\mathrm{G})}_{\mathrm{A}}\right)}{c^2}\right]\left(\mathbf{r}^{(\mathrm{B})}\cdot\mathbf{v}^{(\mathrm{B})}_{\mathrm{A}}\right)
\end{equation}
from Eq.~\eqref{rGvGA in A} and Eq.~\eqref{rGvGB in A}.

Transposing  G\(\leftrightarrow\)B turns Eq.~\eqref{rGvGB in A}  into
\[
\left(\mathbf{v}^{(\mathrm{B})}_{\mathrm{G}}\cdot\mathbf{r}^{(\mathrm{B})}\right)=
\]
\begin{equation}\label{rBvBG in A}
=\gamma_{ \mathrm{AB}}\left(\mathbf{v}^{(\mathrm{B})}_{\mathrm{G}}\cdot\mathbf{v}^{(\mathrm{B})}_{\mathrm{A}}\right)t^{(\mathrm{A})}+\gamma_{ \mathrm{AB}}\left[1-\frac{\left(\mathbf{v}^{(\mathrm{B})}_{\mathrm{G}}\cdot\mathbf{v}^{(\mathrm{B})}_{\mathrm{A}}\right)}{c^2}\right]\left(\mathbf{v}^{(\mathrm{A})}_{\mathrm{G}}\cdot\mathbf{r}^{(\mathrm{A})}\right)-\gamma_{ \mathrm{AB}}\left(\mathbf{v}^{(\mathrm{A})}_{\mathrm{B}}\cdot\mathbf{r}^{(\mathrm{A})}\right).
\end{equation}

\subsection{Transitivity for the time transformation rule}\label{time transitivity}
To arrive at \eqref{time relation AB} one should simply combine \eqref{time relation BG} with \eqref{time relation AG} and then apply \eqref{rGvGB in A}:
\[
t^{(\mathrm{B})}=\gamma_{ \mathrm{BG}}\left[\gamma_{ \mathrm{AG}}\left[t^{(\mathrm{A})}-\frac{\left(\mathbf{v}^{(\mathrm{A})}_{\mathrm{G}}\cdot\mathbf{r}^{(\mathrm{A})}\right)}{c^2}\right]-\frac{\left(\mathbf{v}^{(\mathrm{G})}_{\mathrm{B}}\cdot\mathbf{r}^{(\mathrm{G})}\right)}{c^2}\right]=
\]
\[
=\gamma_{ \mathrm{BG}}\gamma_{\mathrm{AG}}t^{(\mathrm{A})}- \gamma_{ \mathrm{BG}} \frac{ \gamma_{ \mathrm{AG}}\left(\mathbf{v}^{(\mathrm{A})}_{\mathrm{G}}\cdot\mathbf{r}^{(\mathrm{A})}\right)+\left(\mathbf{v}^{(\mathrm{G})}_{\mathrm{B}}\cdot\mathbf{r}^{(\mathrm{G})}\right) }{c^2}=
\]
\[
=\gamma_{ \mathrm{BG}}\gamma_{ \mathrm{AG}}t^{(\mathrm{A})}-\gamma_{ \mathrm{BG}}\gamma _{\mathrm{AG}}\frac{\left(\mathbf{v}^{(\mathrm{G})}_{\mathrm{B}}\cdot\mathbf{v}^{(\mathrm{G})}_{\mathrm{A}}\right)}{c^2}t^{(\mathrm{A})}-\frac{\gamma_{ \mathrm{BG}}}{c^2}\gamma_{ \mathrm{AG}}\left[1-\frac{\left(\mathbf{v}^{(\mathrm{G})}_{\mathrm{B}}\cdot\mathbf{v}^{(\mathrm{G})}_{\mathrm{A}}\right)}{c^2}\right]\left(\mathbf{r}^{(\mathrm{A})}\cdot\mathbf{v}^{(\mathrm{A})}_{\mathrm{B}}\right)=
\]
\[
=\gamma_{ \mathrm{AB}}t^{(\mathrm{A})}-\gamma_{ \mathrm{AB}}\frac{\mathbf{r}^{(\mathrm{A})}\cdot\mathbf{v}^{(\mathrm{A})}_{\mathrm{B}}}{c^2}.
\]
Here Eq.~\eqref{gamma AB} is also used.

\subsection{Transitivity for the longitudinal length transformation rule}\label{longitudinal transitivity}
To obtain \eqref{longitudinal relation AB} one can combine Eq.~\eqref{rBvBA in G} with the transformation rules \eqref{longitudinal relation AG} and \eqref{time relation AG}:
\[
\left(\mathbf{v}^{(\mathrm{B})}_{\mathrm{A}}\cdot\mathbf{r}^{(\mathrm{B})}\right) =\left(\mathbf{v}^{(\mathrm{G})}_{\mathrm{A}}\cdot\mathbf{r}^{(\mathrm{G})}\right)\gamma_{ \mathrm{BG}}\left[1-\frac{\left(\mathbf{v}^{(\mathrm{B})}_{\mathrm{G}}\cdot\mathbf{v}^{(\mathrm{B})}_{\mathrm{A}}\right)}{c^2}\right]-\gamma _{\mathrm{BG}} \left(\mathbf{v}^{(\mathrm{G})}_{\mathrm{B}}\cdot\mathbf{r}^{(\mathrm{G})}\right)+
\]
 \[
+\gamma _{\mathrm{BG}}\left(\mathbf{v}^{(\mathrm{B})}_{\mathrm{A}}\cdot\mathbf{v}^{(\mathrm{B})}_{\mathrm{G}}\right)t^{(\mathrm{G})}=
-\gamma _{\mathrm{AG}}\left[ \left(\mathbf{v}^{(\mathrm{A})}_{\mathrm{G}}\cdot\mathbf{r}^{(\mathrm{A})}\right)-v_{ \mathrm{AG}}^2t^{(\mathrm{A})}\right] \gamma_{ \mathrm{BG}}\left[1-\frac{\left(\mathbf{v}^{(\mathrm{B})}_{\mathrm{G}}\cdot\mathbf{v}^{(\mathrm{B})}_{\mathrm{A}}\right)}{c^2}\right]-
\]
\[
-\gamma _{\mathrm{BG}}\left[ \gamma _{\mathrm{AG}}\left(\mathbf{v}^{(\mathrm{G})}_{\mathrm{B}}\cdot\mathbf{v}^{(\mathrm{G})}_{\mathrm{A}}\right)t^{(\mathrm{A})}-\gamma_{ \mathrm{AG}}\left(\mathbf{v}^{(\mathrm{A})}_{\mathrm{G}}\cdot\mathbf{r}^{(\mathrm{A})}\right)+\gamma_{ \mathrm{AG}}\left[1-\frac{\left(\mathbf{v}^{(\mathrm{G})}_{\mathrm{B}}\cdot\mathbf{v}^{(\mathrm{G})}_{\mathrm{A}}\right)}{c^2}\right]\left(\mathbf{r}^{(\mathrm{A})}\cdot\mathbf{v}^{(\mathrm{A})}_{\mathrm{B}}\right)\right]+
\]
\[
+\gamma _{\mathrm{BG}}\left(\mathbf{v}^{(\mathrm{B})}_{\mathrm{A}}\cdot\mathbf{v}^{(\mathrm{B})}_{\mathrm{G}}\right)\gamma_{ \mathrm{AG}}\left[t^{(\mathrm{A})}-\frac{\left(\mathbf{v}^{(\mathrm{A})}_{\mathrm{G}}\cdot\mathbf{r}^{(\mathrm{A})}\right)}{c^2}\right]=-\gamma _{\mathrm{BG}}\gamma_{ \mathrm{AG}}\left[1-\frac{\left(\mathbf{v}^{(\mathrm{G})}_{\mathrm{B}}\cdot\mathbf{v}^{(\mathrm{G})}_{\mathrm{A}}\right)}{c^2}\right]\left(\mathbf{r}^{(\mathrm{A})}\cdot\mathbf{v}^{(\mathrm{A})}_{\mathrm{B}}\right)+
\]
\[
+t^{(\mathrm{A})}\gamma _{\mathrm{BG}}\gamma_{\mathrm{AG}}\left[\left[1-\frac{\left(\mathbf{v}^{(\mathrm{B})}_{\mathrm{G}}\cdot\mathbf{v}^{(\mathrm{B})}_{\mathrm{A}}\right)}{c^2}\right]v_{ \mathrm{AG}}^2-\left(\mathbf{v}^{(\mathrm{G})}_{\mathrm{B}}\cdot\mathbf{v}^{(\mathrm{G})}_{\mathrm{A}}\right)+\left(\mathbf{v}^{(\mathrm{B})}_{\mathrm{A}}\cdot\mathbf{v}^{(\mathrm{B})}_{\mathrm{G}}\right)\right]=
\]
\[
=-\gamma_{\mathrm{AB}} \left(\mathbf{r}^{(\mathrm{A})}\cdot\mathbf{v}^{(\mathrm{A})}_{\mathrm{B}}\right)+t^{(\mathrm{A})} \gamma_{\mathrm{AB}}v_{ \mathrm{AB}}^2
\]
The last equation uses Eq.~\eqref{gamma AB} and Eq.~\eqref{transposition symmetry BB-GG} rewritten as
\[
\left[1-\frac{\left(\mathbf{v}^{(\mathrm{B})}_{\mathrm{G}}\cdot\mathbf{v}^{(\mathrm{B})}_{\mathrm{A}}\right)}{c^2}\right]v_{ \mathrm{AG}}^2+\left(\mathbf{v}^{(\mathrm{B})}_{\mathrm{G}}\cdot\mathbf{v}^{(\mathrm{B})}_{\mathrm{A}}\right)= \left[1-\frac{\left(\mathbf{v}^{(\mathrm{G})}_{\mathrm{B}}\cdot\mathbf{v}^{(\mathrm{G})}_{\mathrm{A}}\right)}{c^2}\right]v_{ \mathrm{AB}}^2+\left(\mathbf{v}^{(\mathrm{G})}_{\mathrm{B}}\cdot\mathbf{v}^{(\mathrm{G})}_{\mathrm{A}}\right).
\]

\subsection{Transitivity for the transversal relation}\label{transversal transitivity}
It remains to show that Eq.~\eqref{transversal relation AB} follows the transformations \eqref{time relation BG}-\eqref{transversal relation BG} and \eqref{time relation AG}-\eqref{transversal relation AG}.

Let us consider the expression
\[
\mathbf{E}^{(\mathrm{G})}=\mathbf{r}^{(\mathrm{G})}+\tau_{\mathrm{A}}\mathbf{v}^{(\mathrm{G})}_{\mathrm{A}}+\tau_{\mathrm{B}}\mathbf{v}^{(\mathrm{G})}_{\mathrm{B}}
\]
where the numbers \(\tau_{\mathrm{A}}\) and \(\tau_{\mathrm{B}}\) secure that
\begin{equation}\label{normal to vGA}
\left(\mathbf{E}^{(\mathrm{G})}\cdot\mathbf{v}^{(\mathrm{G})}_{\mathrm{A}}\right)=0
\end{equation}
and
\begin{equation}\label{normal to vGB}
\left(\mathbf{E}^{(\mathrm{G})}\cdot\mathbf{v}^{(\mathrm{G})}_{\mathrm{B}}\right)=0.
\end{equation}
Solving these equations with
respect to \(\tau_{\mathrm{A}}\) and \(\tau_{\mathrm{B}}\) yields
\[
\tau_{\mathrm{A}}=\frac{\left(\mathbf{r}^{(\mathrm{G})}\cdot\mathbf{v}^{(\mathrm{G})}_{\mathrm{B}}\right) \left(\mathbf{v}^{(\mathrm{G})}_{\mathrm{A}}\cdot\mathbf{v}^{(\mathrm{G})}_{\mathrm{B}}\right)- \left(\mathbf{r}^{(\mathrm{G})}\cdot\mathbf{v}^{(\mathrm{G})}_{\mathrm{A}}\right)v^2_{\mathrm{BG}} }{v^2_{\mathrm{AG}}v^2_{\mathrm{BG}}-\left(\mathbf{v}^{(\mathrm{G})}_{\mathrm{A}}\cdot\mathbf{v}^{(\mathrm{G})}_{\mathrm{B}}\right)^2},
\]
\[
\tau_{\mathrm{B}}=\frac{\left(\mathbf{r}^{(\mathrm{G})}\cdot\mathbf{v}^{(\mathrm{G})}_{\mathrm{A}}\right) \left(\mathbf{v}^{(\mathrm{G})}_{\mathrm{A}}\cdot\mathbf{v}^{(\mathrm{G})}_{\mathrm{B}}\right)- \left(\mathbf{r}^{(\mathrm{G})}\cdot\mathbf{v}^{(\mathrm{G})}_{\mathrm{B}}\right)v^2_{\mathrm{AG}} }{v^2_{\mathrm{AG}}v^2_{\mathrm{BG}}-\left(\mathbf{v}^{(\mathrm{G})}_{\mathrm{A}}\cdot\mathbf{v}^{(\mathrm{G})}_{\mathrm{B}}\right)^2}.
\]
With an aid of the equations \eqref{rGvGA in A}, \eqref{rGvGB in A}, \eqref {rGvGB in B}, \eqref{rGvGA in B}
 one can re-express \(\tau_{\mathrm{A}}\) and \(\tau_{\mathrm{B}}\) in terms of the variables defined in the spaces A and B:
\begin{equation}\label{alpha in B}
\tau_{\mathrm{A}}=\frac{\gamma _{\mathrm{BG}}\left[1-\frac{\left(\mathbf{v}^{(\mathrm{G})}_{\mathrm{B}}\cdot\mathbf{v}^{(\mathrm{G})}_{\mathrm{A}}\right)}{c^2}\right]\left[\left(\mathbf{v}^{(\mathrm{B})}_{\mathrm{G}}\cdot\mathbf{r}^{(\mathrm{B})}\right)\left(\mathbf{v}^{(\mathrm{B})}_{\mathrm{A}}\cdot\mathbf{v}^{(\mathrm{B})}_{\mathrm{G}}\right)-\left(\mathbf{v}^{(\mathrm{B})}_{\mathrm{A}}\cdot\mathbf{r}^{(\mathrm{B})}\right)v^2_{\mathrm{BG}}\right]}{v^2_{\mathrm{AG}}v^2_{\mathrm{BG}}-\left(\mathbf{v}^{(\mathrm{G})}_{\mathrm{A}}\cdot\mathbf{v}^{(\mathrm{G})}_{\mathrm{B}}\right)^2}
\end{equation}
\begin{equation}\label{beta in A}
\tau_{\mathrm{B}}=\frac{\gamma _{\mathrm{AG}}\left[1-\frac{\left(\mathbf{v}^{(\mathrm{G})}_{\mathrm{B}}\cdot\mathbf{v}^{(\mathrm{G})}_{\mathrm{A}}\right)}{c^2}\right]\left[\left(\mathbf{v}^{(\mathrm{A})}_{\mathrm{G}}\cdot\mathbf{r}^{(\mathrm{A})}\right)\left(\mathbf{v}^{(\mathrm{A})}_{\mathrm{B}}\cdot\mathbf{v}^{(\mathrm{A})}_{\mathrm{G}}\right)-\left(\mathbf{v}^{(\mathrm{A})}_{\mathrm{B}}\cdot\mathbf{r}^{(\mathrm{A})}\right)v^2_{\mathrm{AG}}\right]}{v^2_{\mathrm{AG}}v^2_{\mathrm{BG}}-\left(\mathbf{v}^{(\mathrm{G})}_{\mathrm{A}}\cdot\mathbf{v}^{(\mathrm{G})}_{\mathrm{B}}\right)^2}
\end{equation}
Here the equations \eqref{gamma AG velocities}, \eqref{gamma BG velocities}, \eqref{velocities product A-G}, \eqref{velocities product B-G} are also used.

Due to Eq.~\eqref{normal to vGB} one can write
\[
\mathbf{E}^{(\mathrm{G})}=\mathbf{r}^{(\mathrm{G})}-\frac{\left(\mathbf{r}^{(\mathrm{G})}\cdot\mathbf{v}^{(\mathrm{G})}_{\mathrm{B}}\right)\mathbf{v}^{(\mathrm{G})}_{\mathrm{B}}}{v_{\mathrm{BG}}^2}+\tau_{\mathrm{A}}\left[\mathbf{v}^{(\mathrm{G})}_{\mathrm{A}}-\frac{ \left(\mathbf{v}^{(\mathrm{G})}_{\mathrm{A}}\cdot\mathbf{v}^{(\mathrm{G})}_{\mathrm{B}}\right) \mathbf{v}^{(\mathrm{G})}_{\mathrm{B}} }{v_{\mathrm{BG}}^2}\right]\backsim
\]
\[
\backsim \mathbf{r}^{(\mathrm{B})}-\frac{\left(\mathbf{r}^{(\mathrm{B})}\cdot\mathbf{v}^{(\mathrm{B})}_{\mathrm{G}}\right)\mathbf{v}^{(\mathrm{B})}_{\mathrm{G}}}{v_{\mathrm{BG}}^2}+\tau_{\mathrm{A}}\frac{\left[\mathbf{v}^{(\mathrm{B})}_{\mathrm{A}}-\frac{\left(\mathbf{v}^{(\mathrm{B})}_{\mathrm{A}}\cdot\mathbf{v}^{(\mathrm{B})}_{\mathrm{G}}\right)\mathbf{v}^{(\mathrm{B})}_{\mathrm{G}}}{v_{\mathrm{BG}}^2}\right]}{\gamma_{\mathrm{BG}}\left[1-\frac{\left(\mathbf{v}^{(\mathrm{B})}_{\mathrm{G}}\cdot\mathbf{v}^{(\mathrm{B})}_{\mathrm{A}}\right)}{c^2}\right]}=
\]
\[
= \mathbf{r}^{(\mathrm{B})}-\frac{\left(\mathbf{r}^{(\mathrm{B})}\cdot\mathbf{v}^{(\mathrm{B})}_{\mathrm{G}}\right)\mathbf{v}^{(\mathrm{B})}_{\mathrm{G}}}{v_{\mathrm{BG}}^2}+\left[\mathbf{v}^{(\mathrm{B})}_{\mathrm{A}}-\frac{\left(\mathbf{v}^{(\mathrm{B})}_{\mathrm{A}}\cdot\mathbf{v}^{(\mathrm{B})}_{\mathrm{G}}\right)\mathbf{v}^{(\mathrm{B})}_{\mathrm{G}}}{v_{\mathrm{BG}}^2}\right]\times
\]
\[\times
\frac{  \left[\left(\mathbf{v}^{(\mathrm{B})}_{\mathrm{G}}\cdot\mathbf{r}^{(\mathrm{B})}\right)\left(\mathbf{v}^{(\mathrm{B})}_{\mathrm{A}}\cdot\mathbf{v}^{(\mathrm{B})}_{\mathrm{G}}\right)-\left(\mathbf{v}^{(\mathrm{B})}_{\mathrm{A}}\cdot\mathbf{r}^{(\mathrm{B})}\right)v^2_{\mathrm{BG}}\right]
}{
v^2_{\mathrm{AB}}v^2_{\mathrm{BG}}-\left(\mathbf{v}^{(\mathrm{B})}_{\mathrm{A}}\cdot\mathbf{v}^{(\mathrm{B})}_{\mathrm{G}}\right)^2}.
\]
The last equation follows Eq.~\eqref{alpha in B} along with Eq.~\eqref{transposition symmetry GG2-BB2}.

The above last expression is perpendicular to \(\mathbf{v}^{(\mathrm{B})}_{\mathrm{A}}\).
To make it evident one needs to re-arrange the terms only. This yields
\begin{equation}\label{E in B}
\mathbf{E}^{(\mathrm{G})}\backsim\mathbf{r}^{(\mathrm{B})}-\frac{\left(\mathbf{r}^{(\mathrm{B})}\cdot\mathbf{v}^{(\mathrm{B})}_{\mathrm{A}}\right)\mathbf{v}^{(\mathrm{B})}_{\mathrm{A}}}{v_{\mathrm{AB}}^2}+\left[\mathbf{v}^{(\mathrm{B})}_{\mathrm{G}}-\frac{\left(\mathbf{v}^{(\mathrm{B})}_{\mathrm{A}}\cdot\mathbf{v}^{(\mathrm{B})}_{\mathrm{G}}\right)\mathbf{v}^{(\mathrm{B})}_{\mathrm{A}}}{v_{\mathrm{AB}}^2}\right]T_{\mathrm{B}}
\end{equation}
where
\begin{equation}\label{TB in B}
T_{\mathrm{B}}=\frac{\left(\mathbf{v}^{(\mathrm{B})}_{\mathrm{A}}\cdot\mathbf{v}^{(\mathrm{B})}_{\mathrm{G}}\right)\left(\mathbf{v}^{(\mathrm{B})}_{\mathrm{A}}\cdot\mathbf{r}^{(\mathrm{B})}\right)-v^2_{\mathrm{AB}}\left(\mathbf{v}^{(\mathrm{B})}_{\mathrm{G}}\cdot\mathbf{r}^{(\mathrm{B})}\right)
}{
v^2_{\mathrm{AB}}v^2_{\mathrm{BG}}-\left(\mathbf{v}^{(\mathrm{B})}_{\mathrm{A}}\cdot\mathbf{v}^{(\mathrm{B})}_{\mathrm{G}}\right)^2}.
\end{equation}

With transposing A\(\leftrightarrow\)B in Eq.~\eqref{E in B}, one can also obtain
\begin{equation}\label{E in A}
\mathbf{E}^{(\mathrm{G})}\backsim\mathbf{r}^{(\mathrm{A})}-\frac{\left(\mathbf{r}^{(\mathrm{A})}\cdot\mathbf{v}^{(\mathrm{A})}_{\mathrm{B}}\right)\mathbf{v}^{(\mathrm{A})}_{\mathrm{B}}}{v_{\mathrm{AB}}^2}+\left[\mathbf{v}^{(\mathrm{A})}_{\mathrm{G}}-\frac{\left(\mathbf{v}^{(\mathrm{A})}_{\mathrm{B}}\cdot\mathbf{v}^{(\mathrm{A})}_{\mathrm{G}}\right)\mathbf{v}^{(\mathrm{A})}_{\mathrm{B}}}{v_{\mathrm{AB}}^2}\right]T_{\mathrm{A}}
\end{equation}
where
\begin{equation}\label{TA in A}
T_{\mathrm{A}}=\frac{ \left(\mathbf{v}^{(\mathrm{A})}_{\mathrm{B}}\cdot\mathbf{v}^{(\mathrm{A})}_{\mathrm{G}}\right)\left(\mathbf{v}^{(\mathrm{A})}_{\mathrm{B}}\cdot\mathbf{r}^{(\mathrm{A})}\right)-v^2_{\mathrm{AB}}\left(\mathbf{v}^{(\mathrm{A})}_{\mathrm{G}}\cdot\mathbf{r}^{(\mathrm{A})}\right)
}{
v^2_{\mathrm{AB}}v^2_{\mathrm{AG}}-\left(\mathbf{v}^{(\mathrm{A})}_{\mathrm{B}}\cdot\mathbf{v}^{(\mathrm{A})}_{\mathrm{G}}\right)^2}.
\end{equation}
 
Meanwhile, Eq.~\eqref{rBvBG in A} and the transformation rule \eqref{longitudinal relation AB}  (aready derived in the previous section), with an aid of Eq.~\eqref{velocities product A-B}, entail
\[
\left(\mathbf{v}^{(\mathrm{B})}_{\mathrm{G}}\cdot\mathbf{v}^{(\mathrm{B})}_{\mathrm{A}}\right) \left(\mathbf{v}^{(\mathrm{B})}_{\mathrm{A}}\cdot\mathbf{r}^{(\mathrm{B})}\right)-v_{ \mathrm{AB}}^2\left(\mathbf{v}^{(\mathrm{B})}_{\mathrm{G}}\cdot\mathbf{r}^{(\mathrm{B})}\right)=
\]

\[
=\gamma_{ \mathrm{AB}}v_{\mathrm{AB}}^2\left\{\left[1-\frac{\left(\mathbf{v}^{(\mathrm{B})}_{\mathrm{G}}\cdot\mathbf{v}^{(\mathrm{B})}_{\mathrm{A}}\right)}{v_{ \mathrm{AB}}^2}\right]\left(\mathbf{v}^{(\mathrm{A})}_{\mathrm{B}}\cdot\mathbf{r}^{(\mathrm{A})}\right)-\left[1-\frac{\left(\mathbf{v}^{(\mathrm{B})}_{\mathrm{G}}\cdot\mathbf{v}^{(\mathrm{B})}_{\mathrm{A}}\right)}{c^2}\right]\left(\mathbf{v}^{(\mathrm{A})}_{\mathrm{G}}\cdot\mathbf{r}^{(\mathrm{A})}\right)\right\}=
\]
\[
=\gamma_{ \mathrm{AB}}\left[1-\frac{\left(\mathbf{v}^{(\mathrm{B})}_{\mathrm{G}}\cdot\mathbf{v}^{(\mathrm{B})}_{\mathrm{A}}\right)}{c^2}\right]\left[\left(\mathbf{v}^{(\mathrm{A})}_{\mathrm{G}}\cdot\mathbf{v}^{(\mathrm{A})}_{\mathrm{B}}\right) \left(\mathbf{v}^{(\mathrm{A})}_{\mathrm{B}}\cdot\mathbf{r}^{(\mathrm{A})}\right) - v_{\mathrm{AB}}^2\left(\mathbf{v}^{(\mathrm{A})}_{\mathrm{G}}\cdot\mathbf{r}^{(\mathrm{A})}\right)\right].
\]
This relationship and Eq.~\eqref{transposition symmetry BB2-AA2} yield the relation
\begin{equation}\label{TBTA}
\frac{T_{\mathrm{B}}}{T_{\mathrm{A}}}=\gamma_{\mathrm{AB}}\left[1-\frac{\left(\mathbf{v}^{(\mathrm{B})}_{\mathrm{G}}\cdot\mathbf{v}^{(\mathrm{B})}_{\mathrm{A}}\right)}{c^2}\right]\frac{\gamma_{\mathrm{BG}}^2}{\gamma_{\mathrm{AG}}^2}
\end{equation}
between the quantities defined by Eq.~\eqref{TB in B} and Eq.~\eqref{TA in A}. Then, due to Eq.~\eqref{transposition symmetry BB-AA} and Eq.~\eqref{velocity transversal relation AGB},  Eq.~\eqref{TBTA} leads us to the relation

 \[
\left[\mathbf{v}^{(\mathrm{B})}_{\mathrm{G}}-\frac{\left(\mathbf{v}^{(\mathrm{B})}_{\mathrm{G}}\cdot\mathbf{v}^{(\mathrm{B})}_{\mathrm{A}}\right)\mathbf{v}^{(\mathrm{B})}_{\mathrm{A}}}{v_{\mathrm{AB}}^2}\right]T_{\mathrm{B}}
\backsim
\left[\mathbf{v}^{(\mathrm{A})}_{\mathrm{G}}-\frac{\left(\mathbf{v}^{(\mathrm{A})}_{\mathrm{G}}\cdot\mathbf{v}^{(\mathrm{A})}_{\mathrm{B}}\right)\mathbf{v}^{(\mathrm{A})}_{\mathrm{B}}}{v_{\mathrm{AB}}^2}\right]T_{\mathrm{A}},
\]
which, due to Eq.~\eqref{E in B} and Eq.~\eqref{E in A} and the property \eqref{addition property} results in desired Eq.~\eqref{transversal relation AB}.

\section{Discussion}\label{discussion}
The proceeding section shows that the transitivity of the transformation \eqref{time relation AB}-\eqref{transversal relation AB} is one of the key points that secure the existence of Minkowski space. The transformation not only needs the conditions discussed in Section~\ref{physical spaces} but also presupposes the existence of particles able to interact with each other and electromagnetic field over a short spatial range only so that the particles' motions along with the acts of such interaction explicitly or implicitly underlie the basic effects of the relativity theory.\cite[Sec.~III]{Wagner2016a}

The above remark suggests that the proof of the existence of Minkowski space in Section~\ref{proof} is essentially based on properties of free motion of point particles. But such motion is highly degenerate: an infinite number of initial positions is possible for one trajectory of for a given velocity vector at the place of a given event. Apparently, universal external action, such as gravity, can lower the degree of this degeneracy or even remove it completely.

Then the generalization of Minkowski space is hardly possible, except in the case of high symmetry, such as a spherically symmetric action of gravity. Even if one succeeds in generalizing the concept of physical space to allow for an arbitrary and evolving spatial geometry, perceived by some set of observers, in order to arrive at the full spacetime one has no choice but to postulate the transitivity of the transformation between two sets of observers, which apparently imposes an unnecessary and non-physical restriction.

In addition, it is becoming increasingly clear that the idea of spacetime is consistent with observational data only in conjunction with forced assumptions such as the presence of a considerable amount of unidentified dark matter/energy\cite{WMAP2013, PlanckCollaboration2013.I, PlanckCollaboration2015.XIII}, admittedly exotic, and/or various gravity modifications\cite{PlanckCollaboration2015.XIV}.
 
Evidently, the reasonable, non-exotic, interpretation of observations needs a theoretical approach as less restricted as possible. Thus, modifying or even relinquishing the concept of spacetime seems quite natural.

\section{Conclusion}
The transformation of Einsteinian time variable and Cartesian coordinates between two inertial reference frames does not make it possible to find out whether Minkowski space exists, unless one resorts to the position vectors.

In contrast, a straightforward calculation shows that the transformation of the time and the position vector of a physical event between two physical spaces establishes an equivalence relation between columns made of the time and the position vector of a given event in each space. This means the existence of Minkowski space and shows that the premises for its proof are the same as for the coordinate-free derivation of basic effects of the special relativity theory: use of Einsteinian time variable and motions of point particles able to interact with each other and electromagnetic field over a short spatial range only.

The high degeneracy of free motions of point particles, together with the intricacy of the above mentioned calculation, suggests that a further generalization of Minkowski space is beyond belief, so that the modification or even the abandonment of the concept of spacetime seems quite natural.


\begin{thebibliography} {5}
\bibitem{partition}Someone may object that the way of partitioning the observers can affect the result. An objection of this kind is the delicate issues, which is beyond the scope of this article.
\end{thebibliography}
\end{document}